\def\ie{{\it i.e.}}

\def\~{{$\tilde{\phantom{a}}$}}

\documentclass [12pt] {article}
\usepackage{epsfig}
\usepackage{color}

\def\thebibliography#1{\section{References}\markboth
 {REFERENCES}{REFERENCES}\list
 {[\arabic{enumi}]}{\settowidth\labelwidth{[#1]}\leftmargin\labelwidth
 \advance\leftmargin\labelsep
 \usecounter{enumi}}
 \def\newblock{\hskip .11em plus .33em minus -.07em}
 \sloppy
 \sfcode`\.=1000\relax}
\def\upcite#1{\raise6pt\hbox{\scriptsize
\cite{#1}}}
  \def\lsim{\mathrel {\vcenter {\baselineskip 0pt \kern 0pt
    \hbox{$<$} \kern 0pt \hbox{$\sim$} }}}
    \def\gsim{\mathrel {\vcenter {\baselineskip 0pt \kern 0pt
    \hbox{$>$} \kern 0pt \hbox{$\sim$} }}}

\def\hline{\noalign{\hrule \vskip2pt}}

\def\|{\ifmmode\Vert\else \char`\|\fi}
\ifx\oldzeta\undefined                          
  \let\oldzeta=\zeta                            
  \def\zzeta{{\raise 2pt\hbox{$\oldzeta$}}}     
  \let\zeta=\zzeta                              
\fi

\ifx\oldchi\undefined                           
  \let\oldchi=\chi                              
  \def\cchi{{\raise 2pt\hbox{$\oldchi$}}}       
  \let\chi=\cchi                                
\fi

\def\frac#1#2{{#1 \over #2}}

\def\half{\ifinner {\scriptstyle {1 \over 2}}
   \else {1 \over 2} \fi}

\def\simge{\mathrel{%
   \rlap{\raise 0.511ex \hbox{$>$}}{\lower 0.511ex \hbox{$\sim$}}}}
\def\simle{\mathrel{
   \rlap{\raise 0.511ex \hbox{$<$}}{\lower 0.511ex \hbox{$\sim$}}}}

\def\buildchar#1#2#3{{\null\!                   
   \mathop#1\limits^{#2}_{#3}                   
   \!\null}}                                    
\def\overcirc#1{\buildchar{#1}{\circ}{}}

\def\slashchar#1{\setbox0=\hbox{$#1$}           
   \dimen0=\wd0                                 
   \setbox1=\hbox{/} \dimen1=\wd1               
   \ifdim\dimen0>\dimen1                        
      \rlap{\hbox to \dimen0{\hfil/\hfil}}      
      #1                                        
   \else                                        
      \rlap{\hbox to \dimen1{\hfil$#1$\hfil}}   
      /                                         
   \fi}                                         %

\def\subrightarrow#1{
  \setbox0=\hbox{
    $\displaystyle\mathop{}
    \limits_{#1}$}
  \dimen0=\wd0
  \advance \dimen0 by .5em
  \mathrel{
    \mathop{\hbox to \dimen0{\rightarrowfill}}
       \limits_{#1}}}                           

\def\overlay#1#2{\ifmmode%
\setbox0=\hbox{$#1$}%
\setbox1=\hbox to\wd0{\hss$#2$\hss}\else%
\setbox0=\hbox{#1}%
\setbox1=\hbox to\wd0{\hss#2\hss}\fi%
#1\hskip-\wd0\box1 }

\def\pmb#1{\leavevmode\setbox0=\hbox{#1}%
\kern-.02em\copy0\kern-\wd0
\kern.04em\copy0\kern-\wd0
\kern-.02em\raise.04em\box0 }

\def\vereq#1#2{\lower3pt\vbox{\baselineskip1.5pt \lineskip1.5pt
\ialign{$\m@th#1\hfill##\hfil$\crcr#2\crcr\sim\crcr}}}

\def\tensor#1{\protect\@ontopof{#1}{\leftrightarrow}{1.15}\mathord{\box2}}
\def\overstar#1{\protect\@ontopof{#1}{\ast}{1.15}\mathord{\box2}}
\def\overdots#1{\protect\@ontopof{#1}{\cdots}{1.0}\mathord{\box2}}
\def\overcirc#1{\protect\@ontopof{#1}{\circ}{1.2}\mathord{\box2}}
\def\loarrow#1{\protect\@ontopof{#1}{\leftarrow}{1.15}\mathord{\box2}}
\def\roarrow#1{\protect\@ontopof{#1}{\rightarrow}{1.15}\mathord{\box2}}

\def\@ontopof#1#2#3{%
{\mathchoice
{\@@ontopof{#1}{#2}{#3}\displaystyle\scriptstyle}%
{\@@ontopof{#1}{#2}{#3}\textstyle\scriptstyle}%
{\@@ontopof{#1}{#2}{#3}\scriptstyle\scriptscriptstyle}%
{\@@ontopof{#1}{#2}{#3}\scriptscriptstyle\scriptscriptstyle}%
}%
}

\def\@@ontopof#1#2#3#4#5{%
\setbox0=\hbox{$#4#1$}%
\setbox1=\hbox{$#5#2$}%
\setbox2=\hbox{}\ht2=\ht0 \dp2=\dp0 %
\ifdim\wd0>\wd1 %
\setbox1=\hbox to\wd0{\hss\box1\hss}%
\mathord{\rlap{\raise#3\ht0\box1}\box0}%
\else   %
\setbox1=\hbox to.9\wd1{\hss\box1\hss}%
\setbox0=\hbox to\wd1{\hss$#4\relax#1$\hss}%
\mathord{\rlap{\copy0}\raise#3\ht0\box1}%
\fi
}%

\def\lambdabar{\protect\@lambdabar}
\def\@lambdabar{%
\relax
\bgroup
\def\@tempa{\hbox{\raise.73\ht0
\hbox to0pt{\kern.25\wd0\vrule width.5\wd0
height.1pt depth.1pt\hss}\box0}}%
\mathchoice{\setbox0\hbox{$\displaystyle\lambda$}\@tempa}%
{\setbox0\hbox{$\textstyle\lambda$}\@tempa}%
{\setbox0\hbox{$\scriptstyle\lambda$}\@tempa}%
{\setbox0\hbox{$\scriptscriptstyle\lambda$}\@tempa}%
\egroup
}

\def\corresponds{{\lower.2ex\hbox{=}}{\rm\kern-.75em^\triangle}}
\def\succsim{\succ\kern-.9em_\sim\kern.3em}
\def\precsim{\prec\kern-1em_\sim\kern.3em}
\def\slantfrac#1#2{\kern1em^{#1}\kern-.3em/\kern-.1em_{#2}}

\begin{document}
                                                                
\begin{center}
{\Large\bf Radial Viscous Flow between Two Parallel Annular Plates}
\\

\medskip

Kirk T.~McDonald
\\
{\sl Joseph Henry Laboratories, Princeton University, Princeton, NJ 08544}
\\
(June 25, 2000)
\end{center}

\section{Problem}

Deduce the velocity distribution of steady flow of an incompressible  fluid 
of density $\rho$ and viscosity
$\eta$ between two parallel, coaxial annular plates of inner radii $r_1$, outer
radii $r_2$ and separation $h$ when pressure difference $\Delta P$ is applied
between the inner and outer radii.

As an exact solution of the Navier-Stokes equation appears to be difficult,
it suffices to give an approximate solution assuming that the velocity is 
purely radial, ${\bf v} = v(r,z)\hat {\bf r}$, in a cylindrical coordinate
system $(r,\phi,z)$ whose $z$ axis coincides with that of the two annuli.
Deduce a condition for validity of the approximation.

This problem arises, for example, in considerations of a rotary joint between
two sections of a pipe.  Here, we ignore the extra complication of the
effect of the rotation of one of the annuli on the fluid flow.

\section{Solution}

For an incompressible fluid, the velocity distribution obeys the continuity 
equation
\begin{equation}
\nabla \cdot {\bf v} = 0,
\label{e1}
\end{equation}
in which case
the Navier-Stokes equation for steady, viscous flow is 
\begin{equation}
\rho ({\bf v} \cdot \nabla) {\bf v} = - \nabla P + \eta \nabla^2 {\bf v}.
\label{e2}
\end{equation}
There are only three examples in which analytic solutions to this equation have
been obtained when the nonlinear term $({\bf v} \cdot \nabla) {\bf v}$ is
nonvanishing \cite{Landau}.  

We first review the simpler case of two-dimensional flow between parallel 
plates in sec.~2.1, and then take up the case of radial flow in sec.~2.2.
We will find an analytic solution to the nonlinear Navier-Stokes equation
(\ref{e2}) for radial flow, but this solution cannot satisfy the
the boundary conditions,
\begin{equation}
{\bf v}(z=0) = 0 = {\bf v}(z=h),
\label{e6}
\end{equation}
that the flow velocity vanish next to the plates.
However, in the linear approximation to eq.~(\ref{e2}) we obtain an analytic
form for the radial flow between two annular plates.

\subsection{Two-Dimensional Flow between Parallel Plates}

For guidance, 
we recall that an analytic solution is readily obtained for the related
problem of two-dimensional viscous flow between two parallel plates.  
For example, suppose
that the plates are at the planes $z = 0$ and $z = h$, and that the flow is in 
the $x$ direction, \ie, ${\bf v} = v(x,z) \hat{\bf x}$.
The equation of continuity (\ref{e1}) then tells us that 
$\partial v / \partial x
= 0$, so that the velocity is a function of $z$ only,
\begin{equation}
{\bf v} = v(z) \hat{\bf x}.
\label{e3}
\end{equation}
The $z$ component of the Navier-Stokes equation (\ref{e2}) reduces to
$\partial P / \partial z = 0$, so that the pressure is a function of $x$ only.
The $x$ component of eq.~(\ref{e2}) is
\begin{equation}
{\partial P(x) \over \partial x} = \eta {\partial^2 v(z) \over \partial z^2}.
\label{e4}
\end{equation}
Since the lefthand side is a function of $x$, and the righthand side is a
function of $z$, equation (\ref{e4})  can be satisfied only if both sides are 
constant.  Supposing that the pressure decreases with increasing $x$, 
we write 
\begin{equation}
- {\partial P \over \partial x} = \mbox{constant}
=  {\Delta P \over \Delta x} > 0.
\label{e5}
\end{equation}
Using the boundary conditions (\ref{e6}),
we quickly find that
\begin{equation}
v(z) = {\Delta P \over \Delta x} {z (h - z) \over 2 \eta} 
= 6 \bar v {z \over h} \left( 1 - {z \over h} \right),
\label{e7}
\end{equation}
where the average velocity $\bar v$ is given by
\begin{equation}
\bar v = {1 \over h} \int_0^h v(z) dz 
=  {\Delta P \over \Delta x} {h^2 \over 12 \eta}.
\label{e8}
\end{equation}

\subsection{Radial Flow between Parallel Annular Plates}

Returning to the problem of radial flow between two annular plates, we seek
a solution in which the velocity is purely radial,
${\bf v} = v(r,z) \hat{\bf r}$.
The continuity equation (\ref{e1}) for this hypothesis tells us that
\begin{equation}
{1 \over r} {\partial (rv) \over \partial r} = 0,
\label{e9}
\end{equation}
so that
\begin{equation}
{\bf v} = { f(z) \over r} \hat{\bf r}.
\label{e10}
\end{equation}
Following the example of two-dimensional flow between parallel plates, we expect
a parabolic profile in $z$ as in eq.~(\ref{e7}),
\begin{equation}
f(z) \propto z (h - z),
\label{e11}
\end{equation}
which satisfies the boundary conditions (\ref{e6}).

Using the trial solution (\ref{e10}), the $z$ component of the Navier-Stokes
equation (\ref{e2}) again tells us that the pressure must be independent of
$z$: $P = P(r)$.  The radial component of eq.~(\ref{e2}) yields the nonlinear
form 
\begin{equation}
\eta r^2 {d^2 f \over dz^2} + \rho f^2 = r^3 {dP \over dr}.
\label{e12}
\end{equation}

The hoped-for separation of this equation can only be achieved if $f(z) = F$ is
constant, which requires the  pressure profile to be $P(r) = 
A - \rho F^2/2 r^2$.
The boundary conditions (\ref{e6}) cannot be satisfied by this solution.
Further,
this solution exists only for the case that the pressure is increasing with 
increasing radius.  The fluid flow must be then be inward,
so the constant $F$ must be negative. 
The Navier-Stokes equation is not time-reversal invariant
due to the dissipation of energy associated with the viscosity, and
so reversing the velocity of a solution does not, in general, lead to another
solution.
  
While we have
obtained an analytic solution to the nonlinear Navier-Stokes equation 
(\ref{e2}), it is not a solution to the problem of radial flow between two
annuli.  It is hard to imagine a physical problem involving steady, radially 
inward flow of a long tube of fluid, to which the solution could apply.

Instead of an exact solution, we are led to seek an approximate solution
in which the nonlinear term $f^2$ of eq.~(\ref{e12}) can be ignored.  In this
case, the differential equation takes the separable form
\begin{equation}
 {d^2 f \over dz^2} = {r \over \eta} {dP \over dr} = \mbox{constant}.
\label{e13}
\end{equation}
Following eq.~(\ref{e7}) we write the solution for $f$ that satisfies the
boundary conditions (\ref{e6}) as
\begin{equation}
f(z) = 6 \bar f {z \over h} \left( 1 - {z \over h} \right),
\label{e14}
\end{equation}
where $\bar f$ is the average of $f(z)$ over the interval $0 \leq z \leq h$.
The part of eq.~(\ref{e13}) that describes the pressure leads to the
solution
\begin{equation}
P(r) = {P_1 \ln r_2/r + P_2 \ln r/r_1 \over \ln r_2/r_1},
\label{e15}
\end{equation}
where $P_i = P(r_i)$.
Plugging the solutions (\ref{e14}) and (\ref{e15}) back into eq.~(\ref{e13}),
we find that
\begin{equation}
\bar f = {h^2 \Delta P \over 12 \eta \ln r_2/r_1},
\label{e17}
\end{equation}
where $\Delta P = P_1 - P_2$.  Hence, the flow velocity is
\begin{equation}
{\bf v}(r,z) = {z (h - z) \Delta P \over 2 \eta r \ln r_2/r_1}
\hat{\bf r},
\label{e19}
\end{equation}
whose average with respect to $z$ is $\bar v(r) = \bar f/r$.
As with all solutions to the linearized Navier-Stokes equation, the velocity
is independent of the density.

For the approximate solution (\ref{e19}) to
be valid, the term $f^2 \approx \bar f^2$ must be small in eq.~(\ref{e12}), 
which requires
\begin{equation}
{\rho h^4 \Delta P \over 144 \eta^2 r_1^2 \ln r_2/r_1} \ll 1.
\label{e20}
\end{equation}
When this condition is not satisfied, the solution must include
velocity components in the $z$ direction that are significant near the
inner and outer radii, while the flow pattern at intermediate radii could be
reasonably well described by eq.~(\ref{e19}).  

If one of the annuli is rotating at angular velocity $\omega$, the radial
flow velocity should still be given approximately by eq.~(\ref{e19}) so long
as $\omega r_2 \lsim \bar v(r_2) = \bar f/r_2$.



\begin{thebibliography}{99}

\bibitem{Landau}
L.~Landau and E.M.~Lifshitz, 
{\em Fluid Mechanics},
2nd ed.~(Pergamon Press, Oxford, 1987), chap.\ 2.


\end{thebibliography}
\end{document}